# Stress induces remodelling of yeast interaction and co-expression networks

Sonja Lehtinen,[a] Francesc Xavier Marsellach[b], Sandra Codlin[b], Alexander Schmidt[c], Mathieu Clément-Ziza[d], Andreas Beyer[d], Jürg Bähler[b], Christine Orengo[e*], Vera Pancaldi[f*]

[a] CoMPLEX, UCL, Gower Street, London, WC1E 6BT, UK
[b] University College London, Department of Genetics, Evolution and Environment, London, WC1E 6BT, UK; University College London, Cancer Institute, London, WC1E 6BT, UK.
[c] Proteomics Core Facility, Biozentrum, University of Basel, Basel, Switzerland
[d] Cellular Networks and Systems Biology, Biotechnology Center, Technische Universität Dresden, Dresden, Germany
[e] Institute of Structural and Molecular Biology, University College London, Gower Street, London WC1E 6BT, UK, Email: c.orengo@ucl.ac.uk
[f] Structural Biology and BioComputing Program, Spanish National Cancer Research Centre (CNIO), 28029 Madrid, Spain, Email: vpancaldi@cnio.es
* Corresponding authors.

## Abstract

Network analysis provides a powerful framework for the interpretation of genome-wide data. While static network approaches have proved fruitful, there is increasing interest in the insights gained from the analysis of cellular networks under different conditions. In this work, we study the effect of stress on cellular networks in fission yeast. Stress elicits a sophisticated and large scale cellular response, involving a shift of resources from cell growth and metabolism towards protection and maintenance. Previous work has suggested that these changes can be appreciated at the network level. In this paper, we study two types of cellular networks: gene co-regulation networks and weighted protein interaction networks. We show that in response to oxidative stress, the co-regulation networks re-organize towards a more modularised structure: while sets of genes become more tightly co-regulated, co-regulation between these modules is decreased. This shift translates into longer average shortest path length, increased transitivity, and decreased modular overlap in these networks. We also find a similar change in structure in the weighted protein interaction network in response to both oxidative stress and nitrogen starvation, confirming and extending previous findings. These changes in network structure could represent an increase in network robustness and/or the emergence of more specialised functional modules. Additionally, we find stress induces tighter co-regulation of non-coding RNAs, decreased functional importance of splicing factors, as well as changes in the centrality of genes involved in chromatin organization, cytoskeleton organization, cell division, and protein turnover.

## Introduction

Biological systems are endowed with a considerable ability to adapt to different environments, allowing them to survive external insults by launching prompt and pervasive rearrangements of their regulatory systems.[1,2] In the past decades, this phenomenon has been studied extensively based on genome-wide expression analysis, revealing the presence of a large subset of genes that are up- or down-regulated as part of a sophisticated stress response.[3,2,4,5] The stress response leads to a transient arrest in growth, allowing the cell to invest energy in multiple protective mechanisms[6].

Regulation of the stress response occurs at the transcriptional level as well as post-transcriptional and post-translational levels[7]. Generally, stress elicits the activation of a kinase cascade which culminates in the launch of a transcriptional response. Thus, the rapid and transient initial response is followed by long term cellular changes involving down-regulation of metabolism and growth in favour of defence against stress[1]. This response limits the damage inflicted by the present threat but also promotes resilience against further insults[3].

Genes that are induced under stress are more rapidly evolving and characterized by higher variability between cells and conditions[8,9]. This finding suggests that a variable environment favours higher levels of heterogeneity ('bet hedging'), making it more likely for at least part of the population to survive a change in conditions[1]. It is thus possible that the rearrangements produced during the stress response lead the cell into a more plastic state, allowing it to explore larger portions of phenotypic space and ultimately favouring adaptation. Alternatively, one could imagine that stricter control is necessary to ensure survival of the cell in the face of adverse conditions.

While numerous studies have helped characterise specific stress response pathways and mechanisms[10], the integration of these pathways and mechanisms on a whole cell level is less well understood. Network approaches have proved to be a powerful tool for the analysis of genome-wide datasets; they provide insights into cellular behaviour, for example in explaining cellular adaptability[11] and resilience to perturbations[12]. The focus has recently shifted from the characterization of static networks towards understanding their dynamics and control, as well as the topological alteration produced by changes in the environment[13,14]. For example, Gopalacharyulu et al. compiled an integrated budding yeast network combining physical protein interactions with curated pathways and metabolic data[15]. Their analysis showed a dynamic modulation of the system's connectivity in response to oxidative stress. Similar large scale stress-induced topological changes have also been seen in budding yeast transcriptional networks[16].

More recently, Mihalik and Csermely generated differing networks for stressed and non-stressed states by weighting the budding yeast interactome by the abundance of the interacting proteins in each condition. The authors reported a partial disassociation of this network under heat stress[17], with lesser communication between network modules. The authors suggest that this decoupling of modules represents a cellular survival strategy. The pruning of interactions could (i) decrease information flow between modules, thus minimizing the spread of damage[18]; (ii) represent the emergence of more specialized and autonomic functional units; or (iii), in networks where links have a metabolic cost, be the result of energy saving measures.

Mihalik and Csermely's study used mRNA levels as a proxy for protein abundance and focused on two specific datasets, representing the natural and stressed state, collected in two different laboratories. In this study, we perform a similar analysis, but using protein, not mRNA, abundance, measured at several time points after stress induction. We investigate the effects of two distinct stress types and also explore alternative methods of weighting the interactome. Additionally, the analysis is expanded to gene co-regulation networks. This allowed us to gain further insight into stress induced network changes and to probe the role of non-coding RNAs in the stress response. All datasets used were collected in the same laboratory under standardized conditions.

Our results show a clear change in the co-expression networks towards a more modularised structure: genes within network modules become more tightly co-regulated, while co-regulation between modules is decreased. This change in network structure appears to be translated onto the protein-protein interaction network and to take place in response to both oxidative stress and nitrogen starvation.

**Methods**

**Gene co-expression networks**

Gene co-expression networks were constructed using gene expression data from genetic variants exposed to oxidative stress (0.5 mM hydrogen peroxide, $H_2O_2$). Spearman correlation coefficients were computed across the genetic variants for each gene pair, under both stressed and non-stressed conditions. To generate the networks, a specific number of gene pairs with the highest significant ($p<0.05$) correlation coefficients were considered connected, yielding an unweighted network (see supplementary methods for further details). This ensured that networks compared under stressed and non-stressed conditions were of similar size. We verified our results using different numbers of edges and specific correlation coefficient cut-offs. The effect of stress was found to be the same across the networks generated using these different parameters (see Supplementary Tables 1 and 2). Two distinct sets of gene expression data (microarray and RNAseq) were used to generate networks, yielding a total of four co-expression networks (microarray non-stressed, microarray stressed, RNA-seq non-stressed, RNA-seq stressed). Details of the datasets are outline below:

*Microarray*

The networks referred to as microarray co-expression were built from gene expression levels in knock-out mutants (i.e. genetic variants) at 0 and 60 minutes after exposure to 0.5 mM hydrogen peroxide stress. The mutants used in the correlation calculation were *atf31, ppr1, pap1, aft1/pap1, atf1, sty1* and *pmk1*. The choice of mutants was constrained by the availability of expression data under both stressed and non-stressed conditions, collected in the same lab. For more details about the expression data collection, see[9].

*RNA-seq*

The networks referred to as RNA-seq co-expression were constructed from gene expression levels measured by RNA sequencing in the Bähler laboratory from 117 genetic segregants, at 0 and 60 minutes post exposure to 0.5mM hydrogen peroxide stress (manuscritpt in preparation).

**Protein interaction networks**

The physical protein interaction network for *S. pombe* was downloaded from iRefIndex[19], a database consolidating interactions from a number of repositories (BIND, BioGRID, CORUM, DIP, HPRD, IntAct, MINT, MPact, MPPI and OPHID). To capture stress induced changes in the network, we sought to weight the interactions according to an approximation of the probability of their occurrence under specific conditions. We used two distinct approaches.

The first method was to take the product of the abundances of the proteins involved in the interaction. This approximates the probability of the physical interaction occurring in the cell and is similar to the strategy adopted by Mihalik and Csermely[17]. To adjust for the bias against lowly expressed proteins inherent in this method, the approximated probability of interaction (i.e. the product of the abundances) was normalised by the approximated probability under non-stressed conditions. This normalised product was used to weight the interactions. The weights in the non-stressed network thus all become one, whereas the edge weights in the stressed network reflect the ratio of the probabilities of the interaction occurring pre- and post- stress.

The second way of weighting the interactions was to use the correlation coefficient (from the RNA-seq dataset, as this represents correlation across a larger number of genetic variants, thus giving a better estimate of gene co-expression) as weights for the links. Negatively correlated protein pairs were assigned a weight of zero. This too is an approximation of the probability of the interaction occurring in the cell, as proteins both need to be present for the interaction to occur and the presence of the corresponding RNA can be a useful proxy.

*Protein Abundances*

The protein abundance data used in the first method of edge weighting was collected by mass-spectrometry quantification of proteins from wild type fission yeast cells at 0, 5, 10, 15, 30, 50, 60, 90, 120, 180, 240 minutes post exposure to 0.5mM hydrogen peroxide (Papadakis et al., manuscript in preparation).

To investigate the network effects of a different form of stress, we also built weighted abundance networks from protein abundance data from proliferating and quiescent cells. Quiescent cells had undergone 24 hours of nitrogen starvation prior to protein quantification. Protein levels were measured as described in[20].

**Table 1: Summary of different networks used in this paper**

| Network | Method | Analyses |
| --- | --- | --- |
| Microarray Co-expression | Thresholded correlation coefficients across stress related mutants. | Basic network properties, module overlap, GO category analysis. |
| RNA-seq Co-expression | Thresholded correlation coefficients across 117 genetic segregants. | Basic network properties, module overlap, GO category analysis, role of non-coding genes. |
| Abundance- weighted PPI | Link weight is the product of the abundance of the connected proteins, normalized by this product in the non-stressed state. | Module overlap, Betweenness Centrality |
| Co-expression weighted PPI | Correlation coefficients from RNA-seq data assigned as weights for connected proteins. | Module overlap, GO category analysis |

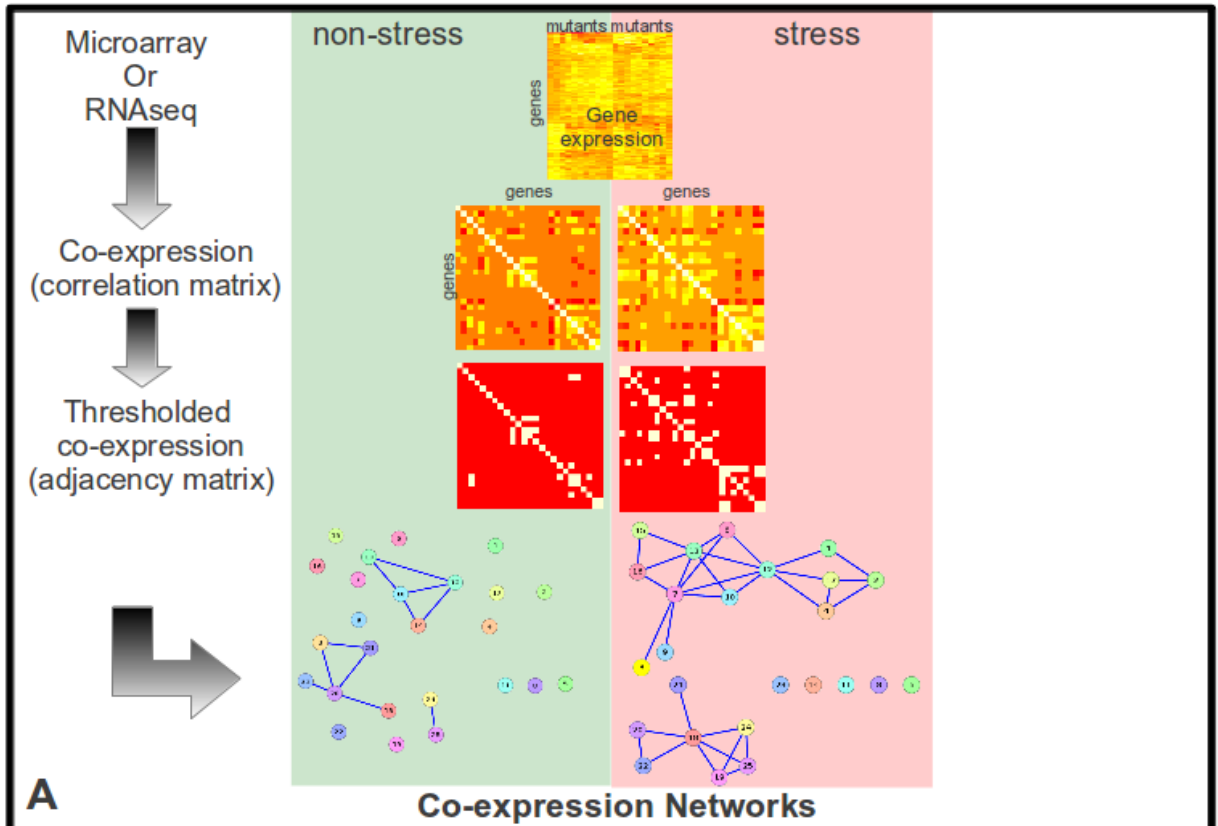

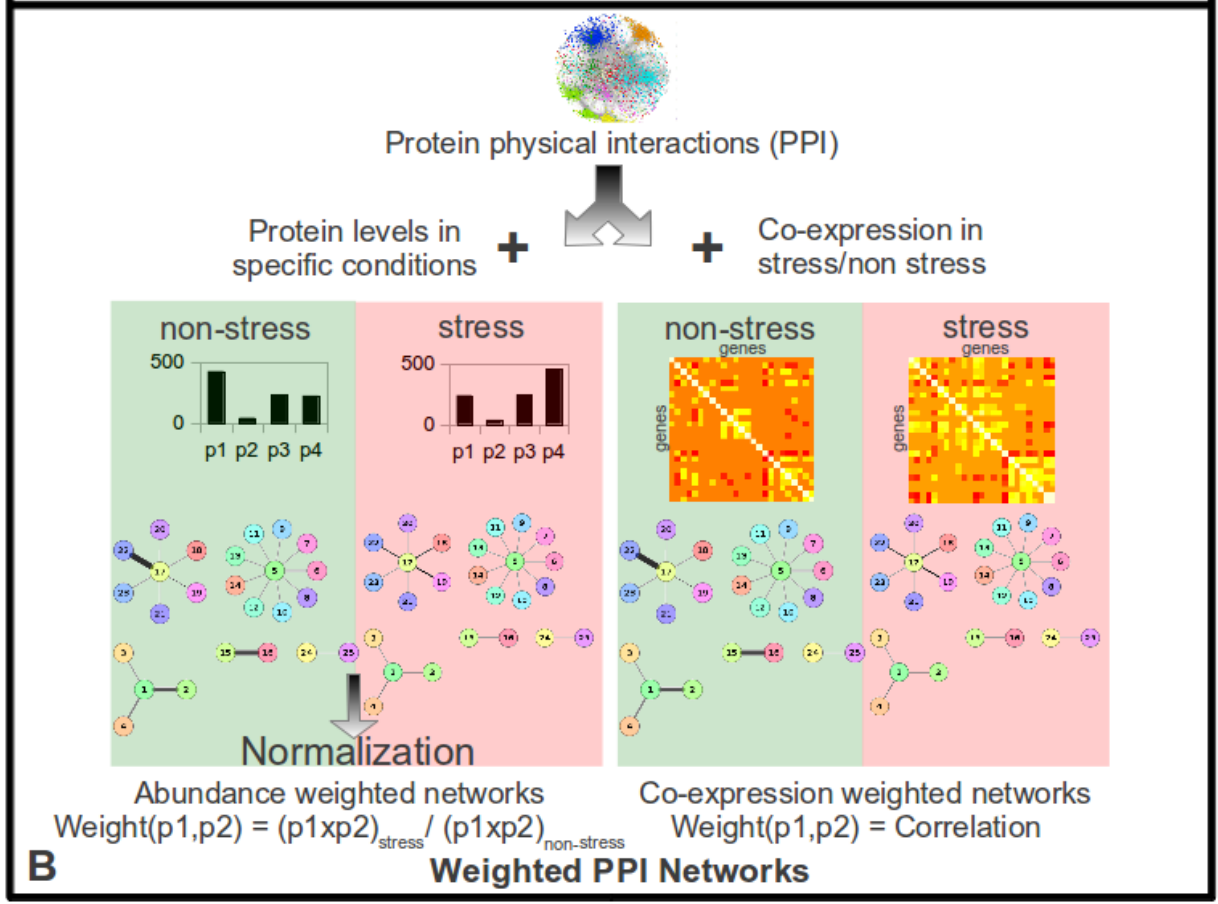

*Figure 1: A general schematic illustrating how the networks used in this paper were generated. A Co-expression networks were generated from gene expression data in various genetic variants under non-stressed and stressed (60 minutes after exposure to 0.5mM hydrogen peroxide) conditions. Correlations in expression between all gene pairs were calculated and the correlation matrix thresholded to give the adjacency matrix of the co-expression network. B: Weighted protein-protein interaction (PPI) networks were generated by condition specific weighting of the physical interaction in fission yeast. The weight of the edge approximates the probability of the interaction occurring in the non-stressed or stressed cell. Two methods of edge weighting were used. 1) Abundance weighting, where the interaction between two proteins was weighted by the product of the proteins' abundances. To avoid bias against lowly expressed proteins, these products were normalized by the product in the non-stressed condition. 2) Co-expression weighting, where the interaction between two proteins was weighted by how correlated their expression is.*

*Modularisation*

Though methods of finding overlapping modules in networks are increasingly popular, no consensus over the best method exists. We therefore used two distinct module finding algorithms: Link Communities[21] and ModuLand[22].

*Link Communities*

Link Communities[21] is based on clustering edges into non-overlapping modules and allowing nodes to inherit all module assignments of their edges, resulting in overlapping module assignment. Similarity, $S$, between edges $e_{ik}$ and $e_{jk}$ was computed as:

$$S(e_{ik}, e_{jk}) = (n_{+(i)} \wedge n_{+(j)}) / (n_{+(i)} \vee n_{+(j)})$$

where $n_+(i)$ is the node i and its neighbours. Edges are assigned into modules by single-linkage hierarchical clustering. For this work, a distance cut-off of 0.4 was used during the hierarchical clustering (see supplementary materials for details on the effect of cut-off). For unweighted networks, the algorithm was implemented using a python script provided by the authors[21]. For weighted networks, the weighted version of the algorithm was implemented with custom written code in *MATLAB*. For unconnected networks, only the largest connected component was considered.

*Moduland*

The ModuLand[22] family of algorithms assigns nodes into modules by computing the *community centrality* of nodes or edges. Community centrality is a measure capturing the influence of nodes or edges on the rest of the network based on a perturbation-flow type calculation. Nodes or edges with higher community centrality than their neighbours are taken as module cores and the other nodes or edges are assigned to modules based on the community centrality values of their neighbours. ModuLand analysis was implemented using the ModuLand Cytoscape plug-in, which uses the LinkLand influence zone determination method and the ProportionalHill module assignment method[23].

*Betweenness Centrality (BC) calculations*
We measured BC in the abundance weighted PPI networks (See supplementary material). To establish the local importance of the nodes, we also measured BCs for specific GO-category sub-networks (BCsub): we constructed 42 sub-networks from the genes belonging to the 42 fission yeast GO Slim terms and measured BC within these sub-networks. Finally, we performed an analysis of linkerity, defined as the ratio of rank of BCsub and rank of BC.[24]

**Results and Discussion**

**Overview of networks**

In order to study stress induced changes on a whole cell level, we constructed two types of network, as illustrated in Figure 1 (see Table 1):
  1. *Gene co-expression networks*: these capture the correlation in gene expression across genetic

variants, thus representing patterns of co-regulation across the genome. We performed our analyses on two distinct networks constructed from different datasets: a microarray dataset of gene expression levels across different mutants and an RNA sequencing datasets across different genetic segregants.

2. *Weighted protein-protein interaction (PPI) networks*: these networks approximate the condition specific probability of physical interactions occurring between proteins. We estimate this probability in two different ways, yielding two distinct networks: firstly using the abundance of the interacting proteins, as measured by mass spectrometry quantification; and secondly using the correlation in gene expression of the interacting proteins.

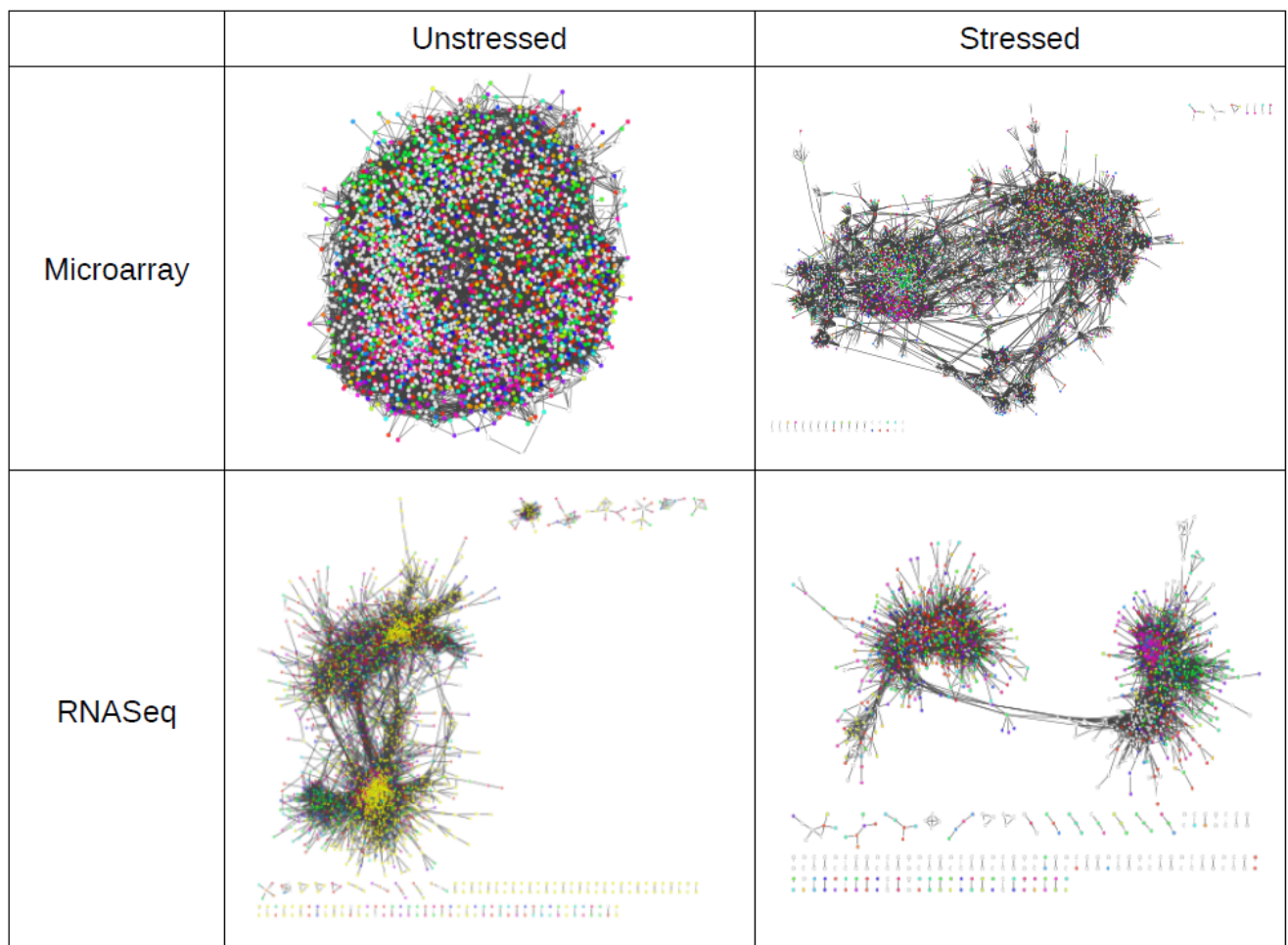

*Figure 2: Visualization of co-expression networks before and after exposure to peroxide stress (0.5mM), showing the re-structuring of the network into more distinct modules. Nodes represent genes while the links between them represent a high level of co-regulation (that is, a high correlation in gene expression across genetic variants). The visualizations were generated using force directed layout in cytoscape and nodes are colour coded according to GO category. Yellow nodes in the RNA-seq unstressed network are either non-coding RNAs or neighbours of a non-coding RNA.*

**Stress Induced Changes in Co-Expression Networks**

*Basic Network Properties: networks become more modularised in response to stress*
We started by constructing gene co-expression networks. The nodes in these networks are genes and the links between them (edges) represent co-regulation. Correlation in gene expression is often one of the main components of functional interaction scores provided by functional network servers, such as String[25]. Given the low coverage and known biases of physical interaction networks, they represent a complementary and distinct perspective. Prior to network generation, we calculated the average of all significant positive correlations under non-stressed and stressed conditions. For both the microarray and RNA-seq datasets, this value was slightly increased (from 0.8698 to 0.8747 and from 0.3216 to 0.3889 respectively), indicating tighter gene co-regulation in response to stress.

The stress induced changes in the co-expression networks can be visually appreciated (Figure 2). However, to meaningfully quantify the stress induced changes in network structure, we computed various network statistics.

We started by calculating the average shortest path length. This is the average minimum number of steps from one node to another in the network, which captures information about the network's connectivity structure. If the network is not fully connected (i.e. paths do not exist between all nodes), only the largest connected component is considered. In both microarray and RNA-seq datasets, stress was found to increase this measure (from 4.58 to 6.10 and from 4.85 to 6.33, respectively). This increase was conserved using different correlation cut-offs for network generation (see Supplementary Table 1).

To check whether this stress induced increase was not simply due to a change in the number of nodes in the network (or in the largest component), we calculated an expected path length for each component size. Twenty control networks were generated for each co-expression network, keeping the same degree structure as the original network but randomly reshuffling the edges (for more details see supplementary materials). Calculating the average shortest path for these control networks gave an *expected* average shortest path length for each network. In both microarray and RNA-seq networks, stress was found to increase the *actual* average shortest path length significantly more than the expected average shortest path length ($p < 10^{-9}$, two-tailed t-test). Therefore, stress causes a restructuring of the network, leading to a longer average shortest path length.

The increase in average shortest path length is particularly noteworthy, given that the density of the largest connected component is increased (from 0.0067 to 0.0068 for the microarray and from 0.012 to 0.026 for the RNA-seq networks). Network density is the number of existing connections divided by the maximum possible number of connections for a fully connected network: a higher network density would thus be expected to yield a shorter path length, as more connections exist in the network. The increase in both path length and density suggests that stress leads to a restructuring of the network where links between "local" genes (i.e. gene pairs that already have short paths between them) are increased, but connections to more "distant" genes become fewer. In other words, the network becomes more modularised.

We tested this idea by looking at the change in transitivity, the likelihood with which two neighbours of a gene are also connected in the network. Consistent with the hypothesis above, stress was found to increase transitivity in both microarray and RNA-seq networks (from 0.38 to 0.42 and 0.52 to 0.60, respectively).

Thus, the increase in path length, transitivity and density all suggest that stress creates a network structure with more tightly co-regulated modules, but fewer inter-modular connections. We further investigated these changes in network structure by directly examining the modular structure of the network.

*Modular Structure: Stress causes a decrease in module overlap in correlation networks*
In biological networks, genes and proteins often participate in more than one function. Because of this feature, the study of *overlapping* modules (that is, allowing a node to belong to multiple network modules) is becoming increasingly popular. However, there is no consensus on the best method to define overlapping modules. Therefore, we decomposed networks into modules using two distinct module finding algorithms:

1. Link Communities (LC): this algorithm divides the links in a network into non-overlapping subsets, which define the link modules. The nodes are then associated with all the modules to which their links belong.
2. ModuLand (ML): this algorithm assigns nodes into modules based on a measure of a node's "influence" on others (see Methods for details).

To understand the stress induced change in the network structure, we looked at changes in module overlap. Module overlap reflects the extent to which a single protein belongs to more than one set of tightly co-regulated proteins. For Link Communities, the overlap is simply the average number of module assignments per node, whereas for ModuLand, the overlap measure takes into account the strength of each module assignment (see Methods).

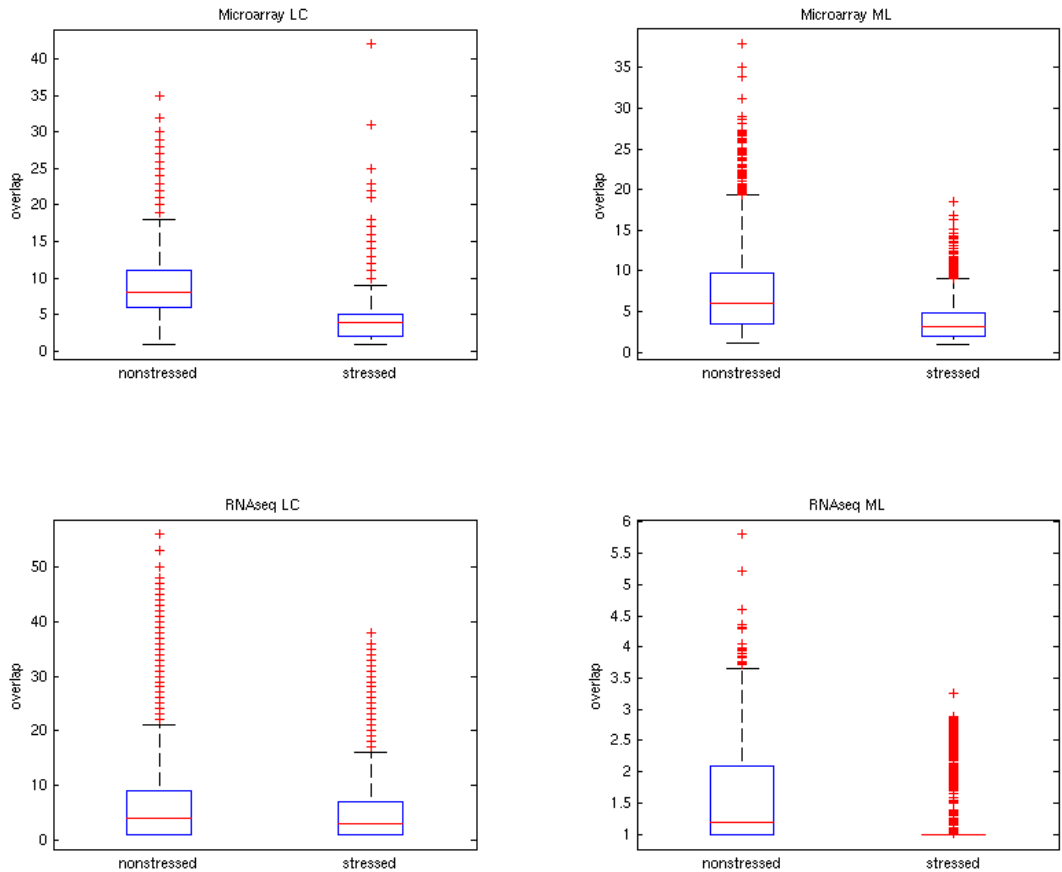

*Figure 3: Changes to modular overlap in co-expression networks in response to oxidative stress (0.5mM hydrogen peroxide). Two distinct module finding algorithms were used: ModuLand (ML) and Link Communities (LC, using clustering cut-off of 0.4, see Methods). For ModuLand modules, overlap was measured as ML overlap (see Methods), while for LC modules, overlap was measured as the number of modules a protein belonged to. Average LC overlap decreased from 8.88 to 3.43 for the microarray network and from 9.98 to 3.31 for the RNAseq network. Average ML overlap decreased from 7.15 to 3.63 for the microarray network and from 1.58 to 1.18 for the RNAseq network. All changes were significant (Wilcoxon ranked sum test, $p<10^{-6}$).*

As seen in Figure 3, overlap decreases significantly in response to stress in both microarray and RNA-seq networks (Wilcoxon ranked sum test, $p<10^{-6}$). This finding is robust when using different thresholds for edge inclusion (Supplementary Table 2). These results confirm the breakdown of the network into modules that have fewer interconnections between them.

This type of structural re-organization is consistent with increased network robustness: decreased communication between functional modules could ensure that perturbations in one module are not spread across the entire network. This increase in robustness could contribute to resilience against further insults.

*Differences between Microarray and RNA-seq networks*
The two co-expression networks were both generated by analysing correlations in gene expression across different genetic variants. The genetic variants in the RNA-seq data are not biased towards stress-related functions and include multiple variations in each strain (derived from crosses of genetically different wild isolates). In the microarray data, on the other hand, all mutants were knock-outs of single genes with known regulatory functions in the stress response.

To test whether the mutant microarray and genetic variant RNA-seq networks capture the same information, we tested the correlation between a gene's co-expression pattern as computed from the two datasets. The average correlation was 0.093 (range: -0.31 to 0.43 Spearman rank correlation).

The low correlation between the two datasets suggests that there is a difference in the information captured by the networks. One explanation for this discrepancy could be that seven genetic conditions are not sufficient to accurately capture gene co-expression. To check whether this was the case, we calculated co-expression using a wider pool of mutants including 24 additional mutants (which could not be used for network construction, because they lacked expression data post exposure to stress). The co-expression, as calculated from these 7 mutants correlated (0.68 Spearman coefficient) with the co-expression as calculated from the 31 mutants. Thus, the 7 mutants are sufficient to produce a fairly representative approximation of co-expression (see Supplementary Information for further discussion of the robustness of the microarray dataset).

A second possible explanation is a bias introduced because all mutants in the microarray dataset are stress related. This could affect the co-expression network in two ways: first, the variability between the genetic conditions is low, explaining the higher average correlation in the microarray dataset. This gives us less power to probe co-expression – some patterns of co-regulation may therefore be missed. Second, the related perturbations could mean that we may not be fully capturing co-regulation for stress related genes: the expression of these genes may be dominated by the direct effects of the perturbation, masking effects of co-regulation.

Despite these points, the stress-induced changes are remarkably consistent in the two networks, suggesting that the effect of stress on the co-expression network is robust.

*Importance of non-coding genes in stress*
We noticed an increase in the presence of non-coding RNAs under stress in the RNA-seq network. The RNA-seq data allowed us to focus the analysis on non-coding RNAs, for which few probes were present in the microarray dataset. Interestingly, non-coding RNAs represent 23% of the set of genes present only in stress, compared to only 13% of genes present only in the non-stressed network and 16% in the genes present in both networks. This finding raises the possibility that the expression of non-coding RNAs becomes more coordinated under stress treatment.

An analysis of the non-coding RNAs that appear to be strongly co-regulated only during stress reveals that the majority are annotated antisense RNAs, overlapping protein coding transcripts on the opposite strand. The corresponding protein-coding transcripts (mRNAs) represent a mixture of cell-cycle factors, chromatin remodellers and metabolism related proteins (data not shown). This suggests these strongly co-regulated non-coding RNAs might play a role in the regulation of these functions during stress. More specifically, we classified links into three classes: links between two non-coding RNAs, links between two coding RNAs and links connecting one coding transcript to a non-coding one. Figure 4 shows the proportion of existing links compared to the total number of possible links within each of these categories, in other words, capturing the density within each of these categories. Stress produces an increase in links connecting the same type of gene (both coding or non-coding) whereas there is no increase in the density of mixed (coding to non-coding) links. This result confirms findings that non-coding antisense RNAs can be regulated independently from their corresponding coding partners.[26] In addition to the antisense RNAs discussed above, some of the non-coding RNAs appearing only in the stressed network are paired with other non-coding RNAs on the opposite strand, while others are intergenic RNAs.

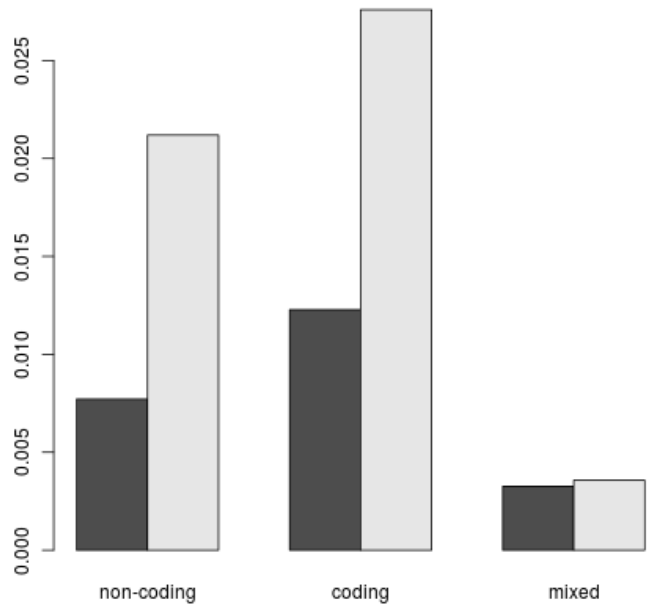

*Figure 4: The density (existing links over possible links) of coding and non-coding RNA sub-networks in the RNAseq co-expression network. The three categories of links shown are: non-coding to non-coding; coding to coding; and non-coding to coding (mixed). Dark bars shows measures for the non-stressed network, lighter bars shows measures for the stressed network. Stress increases the density of coding to coding and non-coding to non-coding links, without greatly affecting the mixed links.*

**Protein Interaction Networks**

*PPI networks show decreased module overlap in response to oxidative stress and nitrogen starvation*
As a complementary approach to the effect of stress on cellular networks, we examined changes in the weighted physical protein interaction (PPI) networks. In these networks, nodes represent proteins while edges are documented physical interactions between them. The same interactome is used for both stress and non-stress, as condition-specific interaction data is not available. To generate condition-specific networks, each edge is assigned a value (weight) representing the probability of this interaction occurring under the specific condition. We used two methods to estimate this probability:

1. Abundance weighting: we approximate the probability of interaction based on the abundance of the interacting proteins. To remove bias against interactions involving lowly expressed proteins, the edge weights were normalised by the edge weight in the non-stressed condition. Because abundance data was available for a limited number of proteins, this network was relatively small (563 proteins).

2. Co-expression weighting: edges in the interactome are weighted according to co-expression; this is also an estimate of the probability of the interaction occurring, as both proteins need to be present at the same time.

The average edge weight using both methods of network construction was increased by stress (from 1 to 2.41 for abundance weighting and from 0.27 to 0.33 for co-expression weighting) indicating that, according to our approximation, the probability of protein-protein interaction is higher after exposure to stress.

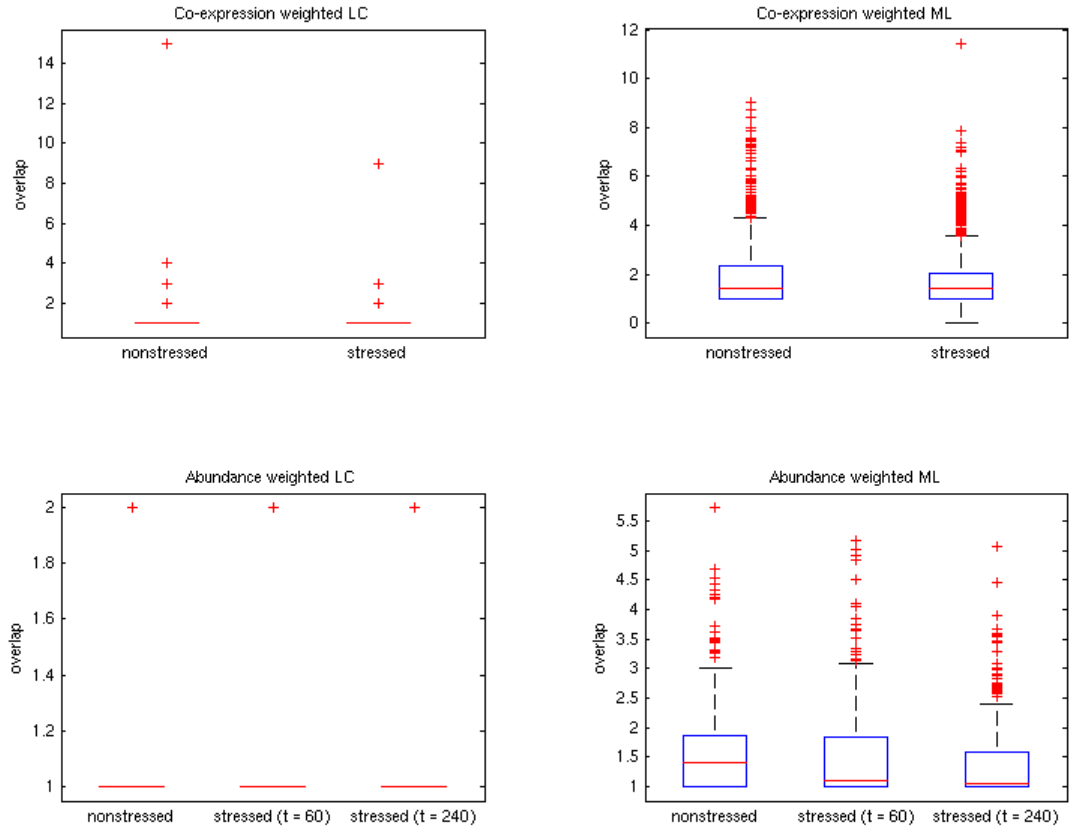

*Figure 5: Changes to modular overlap in response to oxidative stress (0.5mM hydrogen peroxide). The distinct module finding algorithms were used: ModuLand (ML) and Link Communities (LC, using clustering cut-off of 0.4, see Methods). For ModuLand modules, overlap was measured as ML overlap (see Supplementary), while for LC modules, overlap was measured as the number of modules a protein belonged to. Average ML overlap decreases from 1.90 to 1.75 for the co-expression weighted networks and from 1.53 to 1.50 (at t= 60min) and 1.33 (at t= 240min) for the abundance weighted networks. Average LC overlap decreases from 1.12 to 1.04 for co-expression weighting and from 1.0043 to 1.0022 (at t= 60 and 240 min) for abundance weighting. Changes in the ML overlap are significant for the abundance weighted network (Wilcoxon signed rank test, $p<0.001$), though not co-expression weighting ($p = 0.59$).*

We also examined changes in overlap for the PPI networks. Calculating ModuLand overlap confirms a decrease in overlap in response to stress for both methods of networks weighting (Figure 5), though this finding is only significant for the abundance weighting (Wilcoxon signed rank test, $p<0.001$ for abundance weighting, $p = 0.6$ for co-expression weighting). The Link Communities algorithm assigns the vast majority of nodes to a single module (Figure 5) – making analysis of overlap difficult using this algorithm. These effects on the PPI network are less pronounced than in the co-expression network. Although this result may be a genuine difference between the networks, it could also be due to the relatively small coverage of the PPI network, which gives us less statistical power to detect stress induced changes.

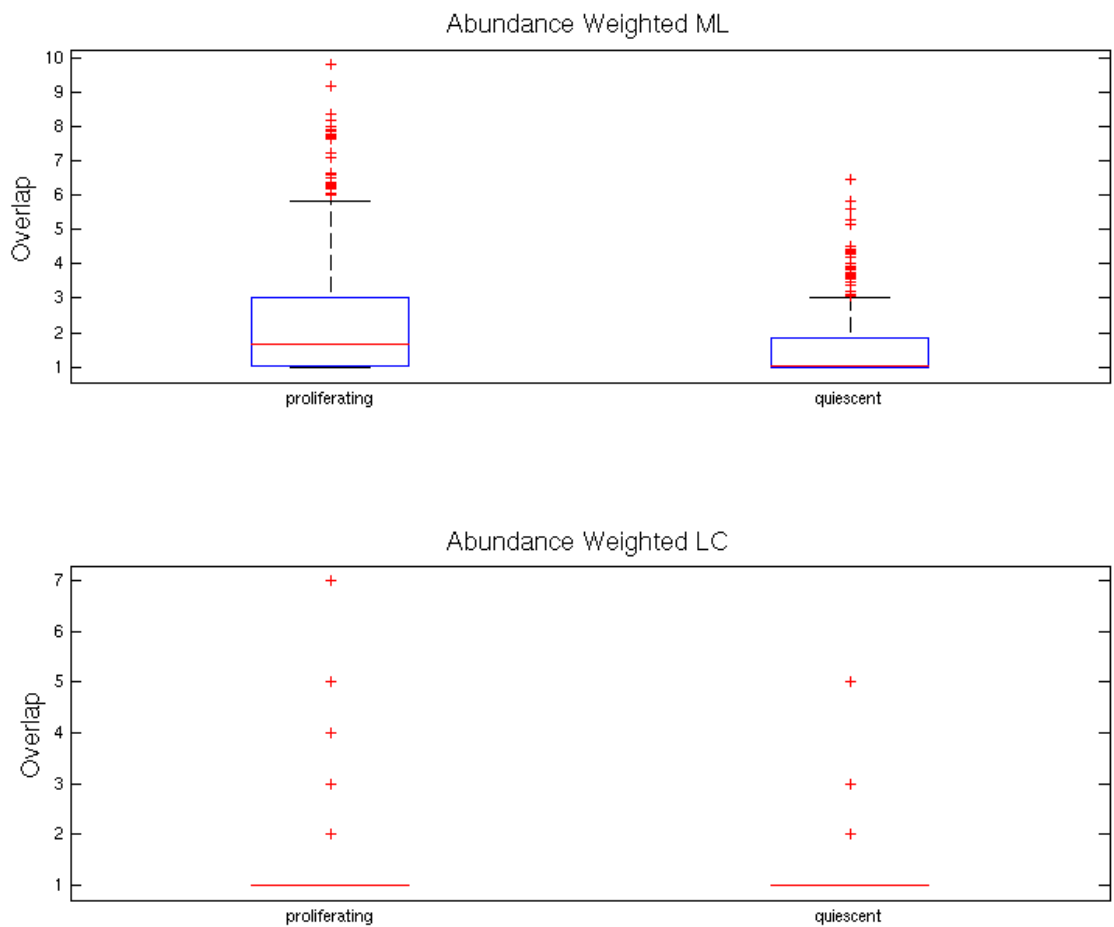

*Figure 6: Changes to modular overlap in proliferating and quiescent cells. Quiescent cells have been exposed to 24 hours of nitrogen starvation. For ModuLand modules, overlap was measured as ML overlap (see Methods), while for LC modules, overlap was measured as the number of modules a protein belonged to. Average ML overlap decreases from 2.23 to 1.53 while LC overlap increases from 1.08 to 1.13. The decrease in ML overlap is significant, Wilcoxon signed rank test, $p<10^{-10}$). Note that these boxplots do not capture the size difference in the networks: therefore, though the proliferation network has nodes with higher LC overlap, its average overlap is lower because of a larger number of nodes with LC overlap of 1.*

To test whether a similar change in network structure is also seen in response to other cellular stresses, we also constructed weighted abundance networks from protein abundance data in response to 24 hours of nitrogen starvation (quiescence). As shown in Figure 6, the ModuLand overlap is also significantly decreased in response to nitrogen starvation (Wilcoxon signed rank test, $p<10^{-10}$). Average Link Communities overlap, however, is increased in response to nitrogen starvation (Wilcoxon signed rank test, $p<10^{-3}$). As with stress, the Link Communities algorithm assigns the majority of the nodes to a single module, again, complicating the interpretation of the results. It is therefore possible that the increase in Link Communities overlap represents a difference in the response to oxidative stress and nitrogen starvation. However, we cannot exclude the possibility that the Link Communities overlap is not adequately capturing the overlap for these networks. It is interesting that, although these two stresses produced different cellular responses, the network effect, as measured by ModuLand overlap, is similar. The potential reasons for the network restructuring – increased robustness, energy saving and development of more distinct functional modules – are plausible responses to both oxidative stress and nitrogen starvation.

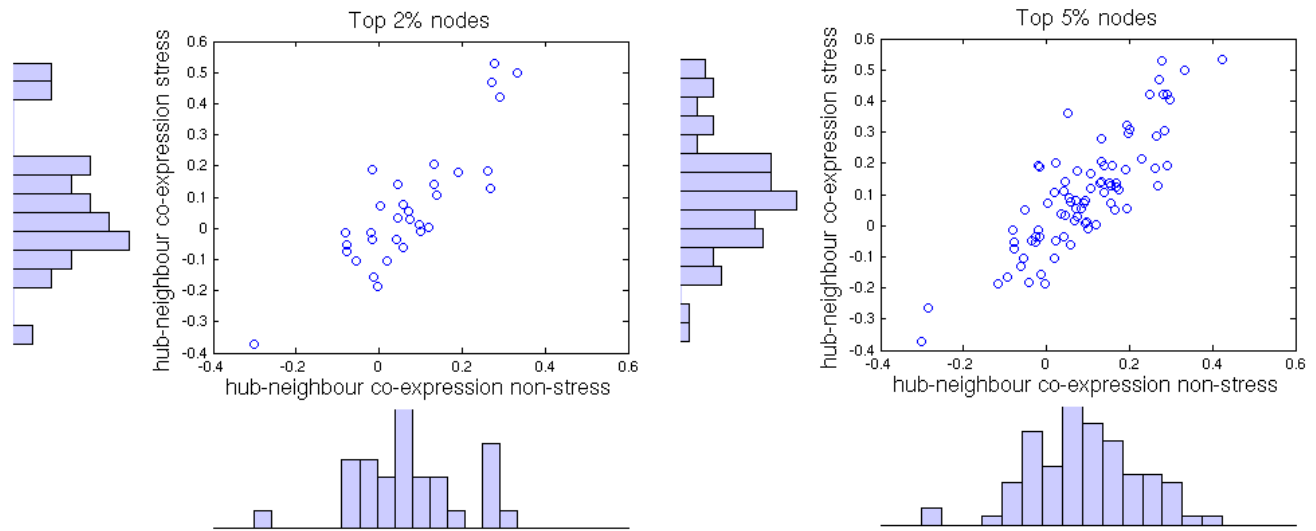

*Figure 7: the effect of stress on the extent to which hubs are co-expressed with their neighbours. Co-expression values represent the average correlation coefficient (calculated from the RNA-seq data) between a hub (top 2% (left) and top 5% (right) most connected nodes) and its neighbours.*

*Further changes in network structure*
To further probe stress induced changes in the network structure, we examined whether there was a change in the extent to which hubs were co-expressed with their neighbours. Figure 7 shows the average co-expression (as calculated from the RNAseq data) between a hub (top 2% and 5% most connected nodes in the PPI network, 32 and 81 proteins respectively) and its neighbours, under both stressed and non- stressed conditions. It appears that the distribution becomes more bimodal with stress: a group of hubs becomes more tightly co-expressed with their neighbours (see supplementary Tables S3 and S4). Given that hubs with little co-expression with their neighbours are thought to function as points of cross-talk between functional modules, the observed change could represent decreased inter-modular communication and the emergence of more tightly co-regulated functional modules. However, given the small size of the PPI network, these conclusions are rather tentative. In particular, we cannot exclude that the change we perceive might represent an extension of the distribution tail instead of bimodality.

*Analysis of node centrality*
Multiple measures can be used to establish the importance of a node. The betweenness centrality (BC) of a node is the proportion of shortest paths in the network which pass through that node. A node with a high BC occupies a central position in the network and a network where most nodes have high BC is likely to be highly interconnected.[27]

We analysed the weighted PPIs (abundance weighted and co-expression weighted) finding that overall neither average BC calculated on the whole network, nor average BC calculated within specific GO-category sub-networks (BCsub) changed significantly after stress treatment. However, when looking at the following sub-networks in the abundance weighted PPIs, we found BCsub to be increased in response to stress: DNA dependent transcription, cytokinesis, nucleocytoplasmic transport and chromatin modification. These functional sub-networks are therefore likely to become more densely interconnected.

Interestingly, BCsub in the chromatin modification sub-network continues to increase at the late stress time point (240 minutes), compared to the 60 minutes point, a behaviour not observed in other GO categories. This is consistent with changes at the chromatin level being part of a more permanent stress induced rearrangement of cellular regulation.[28,29] A more thorough analysis of genes that change BC upon stress treatment is presented in the following section and in supplementary materials (Tables S5, S6, S7 and Figures S4 and S5).

The concept of linkerity expresses the sub-netowrk centrality of a node relative to its global centrality. Nodes with high linkerity values are thus at the edges of sub-networks, but central in the full network. High linkerity proteins are therefore thought to represent linkers between separate functional sub-networks. Although different genes display high linkerity before and after stress treatment, no general change in average linkerity values is observed (see Supplementary Material). However, specific functional categories of genes are seen to change their linkerity values due to stress.

**Biological Correlates of Network Change**

*Mapping network changes onto GO annotations reveals stress induced changes in specific GO categories*
Since we observe changes in network structure in response to stress, we sought to identify which genes and proteins undergo the largest stress-induced change in connectivity. First, we investigated which genes appear in or disappear from the network in response to stress. In the co-expression network, presence of a gene in only the stressed or non-stressed network suggests that it is more tightly co-regulated with other genes in one of the conditions. The RNA-seq co-expression network showed no enrichment for genes present in the stressed network only, while those in the unstressed network only were enriched for ion transmembrane transport and related functions (corrected $p < 10^{-4}$) and regulation of nitrogen compound metabolic processes (corrected $p = 0.005$). Both analyses used the set of genes present in the networks as background to avoid biases towards categories over-represented in the whole network. No enrichment was found in either set of nodes in the microarray, which is not surprising. As discussed previously, there is lesser variability between the genetic variants in this dataset. This leads to a less accurate estimate of gene co-expression, potentially masking some of the stress induced effects on the network.

We performed a similar analysis for PPI networks (only the co-expression weighted, as all genes were present in both networks for abundance weighted networks). Here, the absence of a protein under only one of the conditions is due to all its edges having a weight of zero, indicating that the protein is not important either pre- or post- stress. Neither set of proteins, however, was enriched for any particular GO category when using the PPI network as background.

To investigate changes in connectivity of genes and proteins present in both stressed and unstressed networks, we looked at genes undergoing the largest change in degree following exposure to stress. In the co-expression networks, this indicates which genes' co-expression relations are most affected in response to stress. We sorted the genes according to the magnitude of the change in degree between non-stressed and stressed networks. In the RNA-seq networks, the top 10% of genes with the biggest stress-induced decrease in degree (reaching 173 vs 1) are weakly enriched for monosaccharide catabolic processes (corrected $p=0.0038$). As for the genes with the greatest increase in degree after exposure to stress, the top 10% are enriched for cytoplasmic translation (corrected $p=0.00084$). Again, the genes present in the network were used as the background set for the analysis. In the microarray co-expression network, no enrichment was found in either set of genes, although when using the whole genome as background, the enrichment for cytoplasmic translation in the genes with increasing degree was recovered (corrected $p<10^{-17}$).

We then tested which proteins undergo the largest change in weighted degree in the PPI networks in response to stress. Weighted degree is the sum of the weights of a protein's interaction; thus, in these PPI networks, it represents its probability of participating in an interaction. We examined the 10% of proteins with the greatest stress induced decrease in degree. In the co-expression weighted network, these are enriched (using the rest of the network as background) for mRNA processing and particularly RNA splicing (corrected $p <0.00628$). In the abundance weighted networks, there is no enrichment using the abundance weighted network as background. However, using either the larger PPI network (that is, not excluding proteins for which no proteomics data was available) or the whole genome as background, the mRNA processing and RNA splicing enrichment is recovered (corrected $p<0.00275$). The 10% of proteins undergoing the largest degree increase were not enriched for any GO-terms in either of the networks using the network as background.

In summary, these results suggest a stricter control of proteins involved in translation in the stressed condition. Furthermore, stress appears to decrease the involvement of genes related to RNA splicing in interactions. This finding could reflect that rapidly regulated stress-response genes are under-enriched for introns[30], thus leading to a decreased importance of splicing-related proteins during the stress response. This hypothesis is supported by the finding that the enrichment for splicing related categories is no longer present at 4 hours post exposure to stress.

We repeated the enrichment analysis for sets of proteins undergoing a change in centrality in response to stress. Proteins that decrease their overall BC upon stress treatment are enriched for cytokinesis (corrected $p<10^{-14}$), while the proteins that increase BC are enriched in proteasome subunits (corrected $p<10^{-19}$).

An interesting group of proteins increase their overall BC at the 240 min time-point, and they are enriched in cytoskeleton re-organization (corrected $p<10^{-6}$).

Although the numbers of genes in these lists are small, the enrichments suggest a fundamental role for the proteasome after stress treatment, probably involved in the elimination of the oxidatively damaged protein. Both the enrichment for cytokinesis and cytoskeleton re-organization are likely to be explained by the growth arrest which is initiated during stress response. These findings also suggest an important rearrangement of the cellular structure as a long-term consequence of stress, in line with recent reports of cross-talk between cell cycle and cell shape regulation.[24]

Finally, confirming our previous findings, genes that have changes in linkerity upon stress treatment are enriched for splicing.

## Conclusions

### Gene correlation networks become fragmented in response to stress

Gene correlation networks show higher positive correlation coefficients, longer average shortest path lengths, higher transitivity, and less overlap between modules after exposure to stress. These findings are indicative of a tighter co-regulation between genes within a module, but lesser communication between modules. This type of re-organization might represent the emergence of more specialized functional units in response to stress. It is also consistent with increased network robustness, potentially ensuring resilience to further challenges. Although changes in the weighted PPI networks are more difficult to assess, it appears that the re-organization seen at a gene expression level is indeed translated to the protein level.

Under stress, the co-expression between a group of hubs and their neighbours increases. These findings are reminiscent of a long standing debate about the existence of bimodality in the hub-neighbour co-expression distribution and the distinction between party-hubs (co-expressed with neighbours and binding many partners at once) and date-hubs (not co-expressed with neighbours and binding partners in different places or at different times)[31–33]. We do not believe that our dataset is of a sufficient size to justify any claims in this regard. However, we do observe bimodality in hub-partner co-expression in stressed networks, consistent with the strengthening of inter-module connections parallel to a weakening of intra-module links.

### Specific gene functional categories are seen to be affected by the stress

Our analysis also suggests a decreased importance for splicing factors under stress. This effect is observed in two distinct types of protein interaction network: those weighted according to protein abundance as well as those weighted according to protein co-expression. These genes also present changes in their linkerity upon treatment. The lesser functional importance of this regulatory mechanism after stress exposure could arise from the need for rapid control of genes in response to stress. Importantly, the phenomenon is no longer seen four hours after exposure to stress, highlighting its association with the transient stage of the transcriptional response.

The decreased network centrality of proteins involved in cell division is consistent with the stress-induced growth arrest, while increased centrality of proteasome subunits could indicate a higher turnover of proteins needed to eliminate the oxidatively damaged proteins.

The local betweenness centrality analysis within specific functional subnetworks also suggests the importance of chromatin modification in the long term following stress exposure, accompanied by a rearrangement of the cytoskeleton.

Finally, increased co-expression between non-coding RNAs in the stressed conditions suggests that they might play an important role in cellular stress response.

## Acknowledgements

The authors would like to thank two anonymous referees for suggesting interesting improvements to this work. This research was funded by PhenOxiGEn (an EU FP7 research project) and a Wellcome Trust Senior Investigator Award to JB. S.L. is supported by the Engineering and Physical Sciences Research Council.
A.B. is supported by the Klaus Tschira Foundation. A.B. and M.C.Z are supported by a EU FP7 HEALTH grant (HEALTH-F4-2008-223539).

*Supplementary Material*

**Supplementary Methods**
Co-expression network generation
To generate co-expression networks, we first calculated Spearman correlations in gene expression across different genetic variants (that is, knock-out mutants for the microarray data and genetic segregants for the RNAseq data) for all gene pairs, to generate a correlation matrix C:
$C(i,j) = corr(S_i, S_j)$,
where $S_i$ is a vector containing the expression of gene i in each genetic variant. C can be considered as the adjacency matrix of a network. We then eliminated self-loops ($C(i,i) = 0$) and thresholded C so that it contained either a specific number of edges or only edges above a specific (significant, $p<0.05$) correlation.

**Supplementary Results and Discussion**
*Co-expression network robustness*
Because the co-expression networks built from the microarray data set involved only seven mutants, robustness of the correlation coefficients was verified by sequentially eliminating each mutant from the calculation. For significant correlations above 0.9 this resulted in an average change of 0.02 in magnitude of the correlation coefficients. For significant correlations above 0.7, the change was 0.05. Generating networks from the recalculated correlations resulted in a 0.3% edge gain and 6.75% edge loss when thresholding at 0.9 (gain of 0.3% and loss of 2% when thresholding at 0.7). As a further control, we calculated co-expression using a wider pool of mutants including 24 additional mutants (which could not be used for network construction, because they lacked expression data post exposure to stress). The co-expression, as calculated from the 7 mutants correlated (0.6826 Spearman correlation coefficient) with the co-expression as calculated from the larger set of mutants. These results indicate that the correlation calculation is robust despite the relatively small number of mutants.
Generation of control networks
The variance of the average shortest path lengths (i.e. "expected path lengths") of the 20 control networks was low (standard deviations of the average shortest path length ranged from 0.001 to 0.004 in the 4 sets of 20 control networks). Given this low variability and the computational expense of generating these control networks, we deemed 20 networks to be sufficient to establish the expected path length.
Effect of hierarchical clustering cut-off on Link Communities (LC) overlap
Hierarchical clustering in this paper was performed using a distance cut-off of 0.4. For the co-expression networks, the stress-induced decrease in modular overlap was conserved at other cut-off thresholds (0.3-0.5). For the PPI networks, hierarchical clustering with a threshold of 0.4 assigns the vast majority of nodes to a single module. At lower cut-off values, all edges were assigned into their own module, essentially meaning that the number of modules a node was assigned to was determined by its degree. We were unable to determine a cut-off value for which the clustering did not fall into one of these extremes. It should be noted that full exploration of cut-off values was prohibited by high computational cost.

**Betweenness centrality analysis**
The betweenness centrality (BC) of a node is defined as the proportion of shortest paths in the network which are found to pass through the node:
$BC(v) = \sum_{i}\sum_{j} (g\_ivj / g\_ij,\ i \neq j, i \neq v, j \neq v)$,
where g_ij is the number of shortest paths between nodes i and j, and g_ivj is the number of shortest paths passing through node v. In a weighted network, such as our PPI networks, the shortest path calculation assigns the weight as the cost of travelling each edge. As, in our network, weights represent ease, not cost, of travel, we used 1/weight in determining the shortest paths.
We do not see a significant general change between the values of BC, BCsub or linkerity (as defined in (Vaggi et al., 2012)) for stressed and non-stressed networks (Figure S1).

Vaggi, F., Dodgson, J., Bajpai, A., Chessel, A., Jordán, F., Sato, M., Carazo-Salas, R. E., et al. (2012). Linkers of cell polarity and cell cycle regulation in the fission yeast protein interaction network. *PLoS computational biology*, *8*(10), e1002732. doi:10.1371/journal.pcbi.1002732

**Supplementary Figures:**
Supp. Figure S1: Boxplots showing the distributions of BC values in the three conditions (no stress, stress (60 minutes) and late stress (240 minutes) for abundance weighted PPIs (a), corresponding distributions for local BC measured relative to a single sub-network (b) and distribution of the linkerity values (ratio of local to central BC).

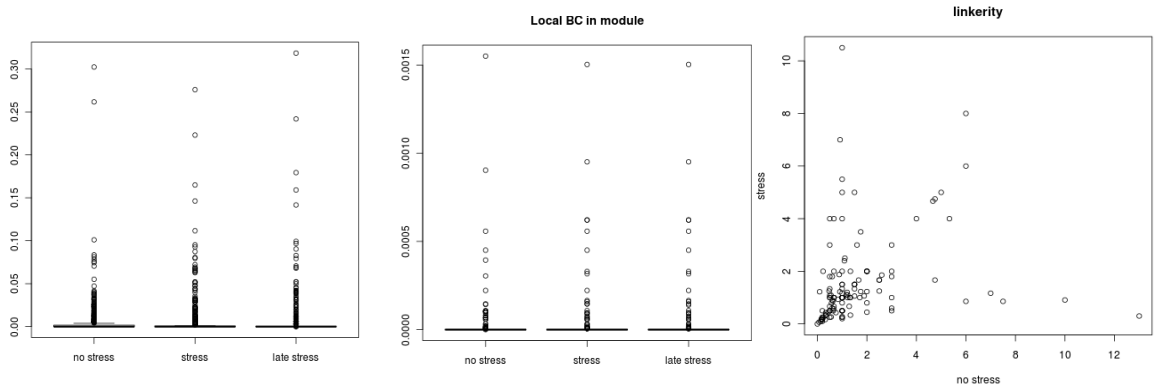

However, we can identify some genes that have higher centrality in the stressed conditions and others that lose centrality with the treatment, especially when considering linkerity (Figure S2).

Supp. Figure S2: Linkerity values for the three difference conditions. Most genes don't change linkerity significantly but there are a few displaying a clear change, meaning that they play a role as a linker between functional modules only in specific conditions.

Some GO category subnetworks show changes in local BC as shown in Figure S3, most changes are driven by single oulying data points. More specific results are shown Table S3 (general statistics), S4 (Top 20 genes that increase BC during stress) and S5 (top genes that increase BC at late stress). Figures S4 and S5 show predicted interactions for these genes. A more thorough analysis of the genes is beyond the scope of this work and will be considered in a future publication.

Supp. Figure S3: Distributions of sub-network specific local BC for the different GO categories for non-stressed and stressed conditions. Very few differences can be seen, mostly due to outliers in the distributions, leading to no general trend.

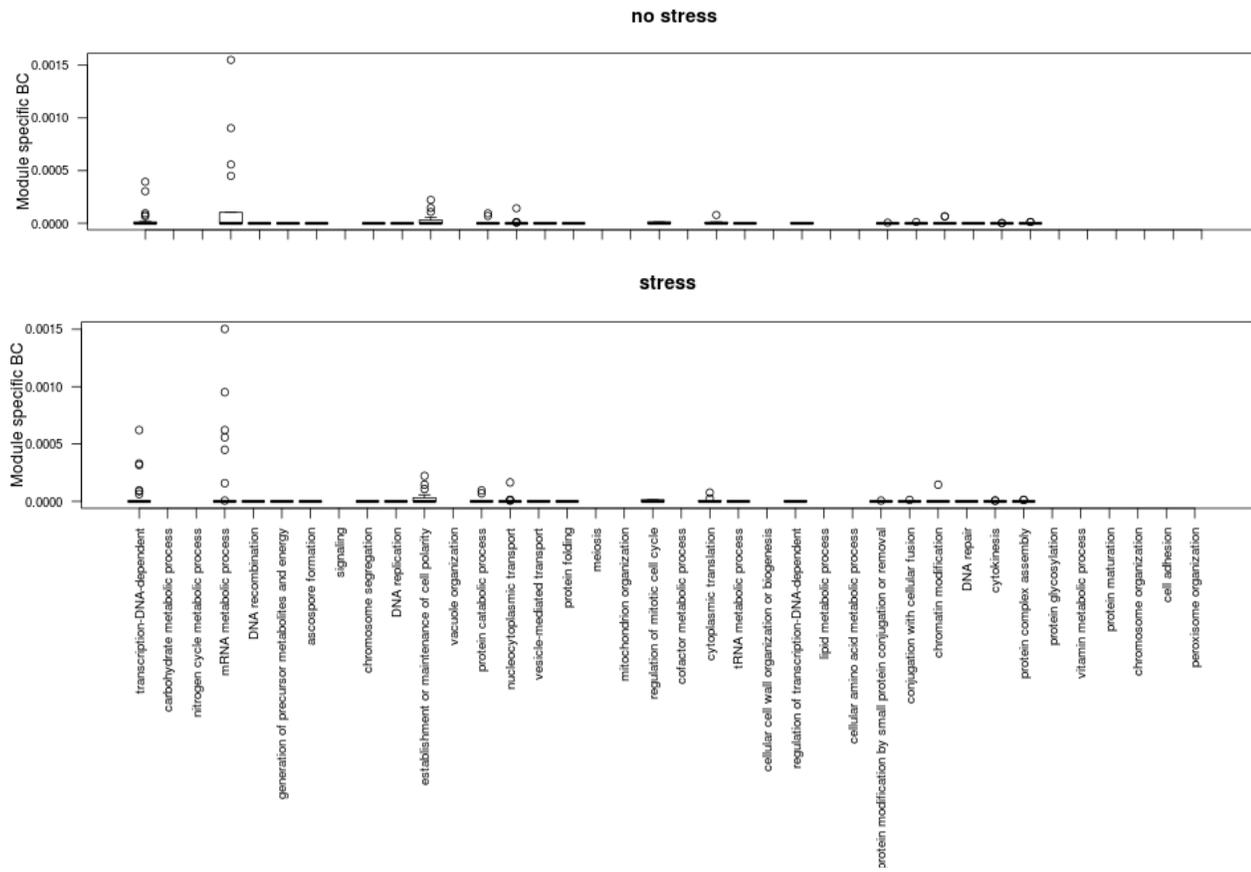

# Supplementary Tables

***Supp Table S1***: *Properties of co-expression networks at various time points in a peroxide stress (0.5mM) time course. Correlations in gene expression were calculated from expression data collected at specific points during the time course in different genetic variants. Networks were generated from the correlation data using two methods: either by drawing connections between a specific number of gene-pairs with the highest correlations, or by considering gene-pairs with a correlation above a specific threshold to be connected. As shown in the table, stress increases network density, shortest average path length and transitivity.*

| Network | Time point (min) | Threshold | Nodes | Edges | Density | Average Shortest Path Length | Actual/Expected | Size of largest Component | Transitivity |
|---|---|---|---|---|---|---|---|---|---|
| RNAseq | 0 | Top 40000 | 2836 | 40000 | 0.005 | 4.85 | 1.87 | 2582 | 0.52397749 |
| RNAseq | 60 | Top 40000 | 1980 | 40000 | 0.010 | 6.33 | 2.71 | 1768 | 0.60193717 |
| Microarray | 0 | Top 60000 | 4240 | 60000 | 0.003 | 4.58 | 1.62 | 4240 | 0.37599285 |
| Microarray | 60 | Top 60000 | 4268 | 60000 | 0.003 | 6.10 | 2.15 | 4213 | 0.41561613 |
| Microarray | 0 | 0.9 | 4241 | 81946 | 0.0091 | 4.43 | 1.68034698 | 4241 | 0.43247021 |
| Microarray | 15 | 0.9 | 4359 | 91406 | 0.0096 | 4.45 | 1.70826335 | 4359 | 0.49322151 |
| Microarray | 60 | 0.9 | 4351 | 186982 | 0.0198 | 4.58 | 1.94748246 | 4351 | 0.56839195 |
| Microarray | 0 | 0.7 | 4242 | 309284 | 0.0344 | 3.20 | 1.55585666 | 4242 | 0.49175417 |
| Microarray | 15 | 0.7 | 4395 | 345603 | 0.0358 | 3.19 | 1.55714104 | 4395 | 0.52946568 |
| Microarray | 60 | 0.7 | 4356 | 619219 | 0.0653 | 3.25 | 1.62504025 | 4356 | 0.60423823 |
| RNAseq | 0 | 0.8 | 832 | 4304 | 0.0125 | 2.49 | 1.11982083 | 259 | 0.54 |
| RNAseq | 60 | 0.8 | 2273 | 54240 | 0.0210 | 8.66 | 1.84501845 | 2075 | 0.6 |

***Supp Table S2***: *Modular properties of co-expression networks, at various points in a peroxide (0.5mM) stress time course, using two different methods of module detection: Moduland (ML) and Link Communities (LC). Using both algorithms, modular overlap (as measured by ModuLand overlap, see Methods, or the average number of Link Communities modules per node) is decreased in response to stress.*

| Dataset | Time Point (minutes) | Threshold | ML Modules | Links between modules (ML) | Module Density (ML) | Average node-wise overlap (ML) | Average modules per node (link communities – S = 0.4) |
|---|---|---|---|---|---|---|---|
| Microarray | 0 | 0.9 | 636 | 200981 | *1.00* | 94.78 | 7.60 |
| Microarray | 15 | 0.9 | 641 | 202952 | *0.99* | 92.36 | 7.70 |
| Microarray | 60 | 0.9 | 439 | 93144 | *0.97* | 42.81 | 5.74 |
| Microarray | 0 | Top 60000 | 458 | 102313 | *0.98* | *7.15* | 8.88 |
| Microarray | 15 | Top 60000 | 551 | 143856 | *0.95* | 7.06 | 6.50 |
| Microarray | 60 | Top 60000 | 339 | 27816 | *0.49* | 3.63 | 3.44 |
| RNAseq | 0 | Top 10000 | 131 | 39 | *0.00* | 1.10 | 3.67 |
| RNAseq | 60 | Top 10000 | 113 | 13 | *0.00* | 1.03 | 3.34 |
| RNAseq | 0 | Top 20000 | 152 | 228 | *0.02* | 1.47 | 4.63 |
| RNAseq | 60 | Top 20000 | 104 | 26 | *0.00* | 1.06 | 4.55 |
| RNAseq | 0 | Top 40000 | 131 | 252 | *0.03* | 1.58 | 9.98 |
| RNAseq | 60 | Top 40000 | 110 | 35 | *0.01* | 1.18 | 6.04 |

**Supp Table S3**: *Hubs in the netwok which decrease co-expression with their neighbours upon stress treatment.*

'SPBP8B7.28c':
  SPBP8B7.28c-1 - Meiotic chromosome segregation protein P8B7.28c; Required for meiotic chromosome segregation

'SPAC29A4.08c':
  cwf8 - Cell cycle control protein cwf8; Involved in mRNA splicing where it associates with cdc5 and the other cwf proteins as part of the spliceosome

'SPBC6B1.10':
  prp17 - Pre-mRNA-processing factor 17; Functions in the second step of pre-mRNA splicing. Involved in splicing intron which are longer than 200 nucleotides

'SPAC644.12':
  cdc5 - Pre-mRNA-splicing factor cef1; Involved in mRNA splicing and cell cycle control

'SPBC15D4.03':
  slm9 - Histone transcription regulator slm9; Probably required for replication-independent chromatin assembly (By similarity). Required for transcriptional silencing in the outer repeat (otr) centromeric repeats and the Tf2 long terminal repeat retrotransposons. May play an indirect role in the regulation of cdc2 and/or wee1 at the G2/M stage of mitosis

'SPAC1782.03':
  SPAC1782.03-1 - Uncharacterized protein C1782.03

'SPBC646.13':
  sds23 - Protein sds23/moc1; Required for normal DNA replication and for proper mitosis. Induces sexual development and ascus formation

'SPAC13C5.02':
  dre4 - DNA replication protein 4

'SPAC821.07c':
  moc3 - Transcriptional regulatory protein moc3; Induces sexual development and ascus formation. Also involved in calcium homeostasis

'SPAC2F3.14c':
  SPAC2F3.14c-1 - WW domain-containing protein C2F3.14c

'SPBC6B1.07':
  prp1 - Pre-mRNA-splicing factor prp1; Involved in pre-mRNA splicing. Interacts with prp6 and prp13. May also be involved in the regulation of the G0-G1/G2 transition. Required for pre-spliceosome formation, which is the first step of pre-mRNA splicing. This protein is associated with snRNP U5. Has a role in branch site-3' splice site selection. Associates with the branch site-3' splice 3'-exon region

'SPBC11B10.09':
  cdc2 - Cell division control protein 2; Plays a key role in the control of the eukaryotic cell cycle. It is required for entry into S-phase and mitosis. When complexed with cig2, plays a role in G1-S phase transition. When activated and complexed with the cyclin cdc13, it leads to the onset of mitosis. p34 is a component of the kinase complex that phosphorylates the repetitive C-terminus of RNA polymerase II. Involved in cell cycle arrest induced by defective RNA splicing. Required for phosphorylation of dis1 to ensure accurate chromosome segregation and for the DNA damage checkpoint

'SPBC947.10':
  SPBC947.10 - Uncharacterized RING finger protein C947.10

'SPCC364.02c':
  bis1 - Stress response protein bis1; Has a role in maintaining cell viability during stationary phase induced by stress response

'SPAC9G1.06c':
  cyk3 - Uncharacterized protein C9G1.06c

'SPBC14F5.08':

med7 - Mediator of RNA polymerase II transcription subunit 7; Component of the Mediator complex, a coactivator involved in the regulated transcription of nearly all RNA polymerase II-dependent genes. Mediator functions as a bridge to convey information from gene-specific regulatory proteins to the basal RNA polymerase II transcription machinery. Mediator is recruited to promoters by direct interactions with regulatory proteins and serves as a scaffold for the assembly of a functional preinitiation complex with RNA polymerase II and the general transcription factors (By similarity)

'SPBC106.04':

ada1 - AMP deaminase; AMP deaminase plays a critical role in energy metabolism

'SPBC83.18c':

SPBC83.18c - C2 domain-containing protein C83.18c

**Supp Table S4**: *Hubs in the netwok which increase co-expression with their neighbours upon stress treatment.*

'SPAC12G12.13c':

cid14 - Poly(A) RNA polymerase cid14; Required for 3' polyadenylation of the 5.8S and 25S rRNAs as a prelude ot their degradation in the exosome. Involved in the nucleolar organization to ensure faithful chromosome segregation during mitosis

'SPAC3A12.11c':

cwf2 - Pre-mRNA-splicing factor cwc2; Involved in pre-mRNA splicing

'SPBP35G2.08c':

air1 - Protein air1; Component of the TRAMP (TRF4) and TRAMP5 complexes which have a poly(A) RNA polymerase activity and are involved in a post- transcriptional quality control mechanism limiting inappropriate expression of genetic information. Polyadenylation is required for the degradative activity of the exosome on several of its nuclear RNA substrates like cryptic transcripts generated by RNA polymerase II and III, or hypomethylated pre-tRNAi-Met. Both complexes polyadenylate RNA processing and degradation intermediates of snRNAs, snoRNAs and mRNAs that accumulate in strains lacking a fun [...]

'SPBC1D7.04':

mlo3 - mRNA export protein mlo3; Has a role in the mRNA export process. Interferes with mitotic chromosome segregation when overexpressed

'SPBC409.05':

skp1 - Suppressor of kinetochore protein 1; Required for cig2 degradation in the G2 and M phases of the cell cycle. Together with pof6, essential for septum processing and cell separation. Involved in mitotic progression, essential for the execution of anaphase B; required for coordinated structural alterations of mitotic spindles and segregation of nuclear membrane structures at anaphase. Involved in the DNA damage checkpoint pathway and maintenance of genome integrity

'SPAC31G5.13':

rpn11 - 26S proteasome regulatory subunit rpn11; Acts as a regulatory subunit of the 26 proteasome which is involved in the ATP-dependent degradation of ubiquitinated proteins

'SPBC409.06':

uch2 - Ubiquitin carboxyl-terminal hydrolase 2; Ubiquitin-protein hydrolase is involved both in the processing of ubiquitin precursors and of ubiquinated proteins. This enzyme is a thiol protease that recognizes and hydrolyzes a peptide bond at the C-terminal glycine of ubiquitin

'SPBC12D12.01':

sad1 - Spindle pole body-associated protein sad1; Associates with the spindle pole body and maintains a functional interface between the nuclear membrane and the microtubule motor proteins. Involved in chromosome segregation during meiosis where it associates with the telomeres

'SPAC8E11.02c':

rad24 - DNA damage checkpoint protein rad24; Required for the DNA damage checkpoint that ensures that DNA damage is repaired before mitosis is attempted. Acts as a negative regulator of meiosis by antagonizing the function of mei2. It inhibits the association of meiRNA (a non-coding RNA molecule required for the nuclear mei2 dot formation) to the phosphorylated but not to the unphosphorylated

    form of mei2 in vitro

'SPAC4F8.13c':

    rng2 - Ras GTPase-activating-like protein rng2; Required for cytokinesis. Component of the contractile F-actin ring; required for its construction following assembly of F-actin at the division site

'SPAC664.01c':

    swi6 - Chromatin-associated protein swi6; Recognizes and binds histone H3 tails methylated at 'Lys-9', leading to epigenetic repression. Involved in the repression of the silent mating-type loci MAT2 and MAT3. May compact MAT2/3 into a heterochromatin-like conformation which represses the transcription of these silent cassettes

'SPBP16F5.03c':

    tra1 - Transcription-associated protein 1; Essential component of histone acetyltransferase (HAT) complexes, which serves as a target for activators during recruitment of HAT complexes. Essential for vegetative growth. Functions as a component of the transcription regulatory histone acetylation (HAT) complexes SAGA, SALSA and SLIK. At the promoters, SAGA is required for recruitment of the basal transcription machinery. It influences RNA polymerase II transcriptional activity through different activities such as TBP interaction and promoter selectivity, interaction with transcription activator [...]

'SPBC28F2.12':

    rpb1 - DNA-directed RNA polymerase II subunit rpb1; DNA-dependent RNA polymerase catalyzes the transcription of DNA into RNA using the four ribonucleoside triphosphates as substrates. Largest and catalytic component of RNA polymerase II which synthesizes mRNA precursors and many functional non-coding RNAs. Forms the polymerase active center together with the second largest subunit. Pol II is the central component of the basal RNA polymerase II transcription machinery. It is composed of mobile elements that move relative to each other. RPB1 is part of the core element with the central large cl [...]

'SPAC1687.20c':

    mis6 - Inner centromere protein mis6; Has a role in the maintenance of core chromatin structure and kinetochore function...

**Supp Table S5**: *Changes in Betweenness Centrality (BC) and subnetwork specific BC (BCsub) statistics in abundance weighted PPIs at the three time points considered. Only GO sub-networks where significant changes were found are shown.*

| Statistic | No stress (t=0) | Stress (60') | Late Stress(240') |
|---|---|---|---|
| Maximum BC value | 0.30 | 0.27 | 0.31 |
| BC standard deviation | 0.02 | 0.02 | 0.02 |
| Mean BCsub transcription DNA dep. | 3.6e-05 | 5.6e-05 | 5.6e-05 |
| Mean BCsub cytokinesis | 4.5e-07 | 9.0e-07 | 9.0e-07 |
| Mean BCsub Nucleocyt. transport | 8.5e-06 | 9.8e-06 | 9.5e-06 |
| Mean BCsub Chrom. Modification | 1.2e-05 | 1.3e-05 | 1.4e-05 |

***Supplementary table S6***: *Genes that have much higher BC after stress (Top 20 genes ranked by fold change)*

'SPCC16C4.18c':
  taf6 - Transcription initiation factor TFIID subunit 6; TAFs are components of the transcription factor IID (TFIID) complex that are essential for mediating regulation of RNA polymerase transcription (By similarity)

'SPBC4B4.03':
  rsc1 - Chromatin structure-remodeling complex subunit rsc1; Component of the chromatin structure remodeling complex (RSC), which is involved in transcription regulation and nucleosome positioning. Controls particularly membrane and organelle development genes

'SPBC3E7.14':
  smf1 - Probable small nuclear ribonucleoprotein F; Probable common Sm protein, is found in U1 and U2 snRNPs (By similarity)

'SPBC16C6.07c':
  rpt1 - 26S protease regulatory subunit 7 homolog; The 26S protease is involved in the ATP-dependent degradation of ubiquitinated proteins. The regulatory (or ATPase) complex confers ATP dependency and substrate specificity to the 26S complex (By similarity)

'SPBC4.07c':
  mts2 - 26S protease regulatory subunit 4 homolog; The 26S protease is involved in the ATP-dependent degradation of ubiquitinated proteins. The regulatory (or ATPase) complex confers ATP dependency and substrate specificity to the 26S complex

'SPBC15D4.14':
  taf73 - Transcription initiation factor TFIID subunit taf73; TAFs are components of the transcription factor IID (TFIID) complex that are essential for mediating regulation of RNA polymerase transcription. Regulates the genes involved in ubiquitin-dependent proteolysis during the progression of M-phase of mitosis

'SPBC839.10':
  SPBC839.10 - Uncharacterized RNA-binding protein C839.10

'SPBC32F12.11':
  SPBC32F12.11-1 - Glyceraldehyde-3-phosphate dehydrogenase 1
  SPBC354.12-1 - Glyceraldehyde-3-phosphate dehydrogenase 2

'SPBC16A3.15c':
  nda2 - Tubulin alpha-1 chain; Tubulin is the major constituent of microtubules. It binds two moles of GTP, one at an exchangeable site on the beta chain and one at a non-exchangeable site on the alpha-chain

'SPBC409.07c':
  wis1 - Protein kinase wis1; Dosage-dependent regulator of mitosis with serine/ threonine protein kinase activity. May play a role in the integration of nutritional sensing with the control over entry into mitosis. It may interact with cdc25, wee1 and win1. May activate sty1

'SPAC19A8.13':
  usp101 - U1 small nuclear ribonucleoprotein 70 kDa homolog; Involved in nuclear mRNA splicing (By similarity). Essential for growth

'SPBC16E9.12c':
  pab2 - Polyadenylate-binding protein 2

'SPAC22H12.04c':
  SPAC13G6.02c-1 - 40S ribosomal protein S3aE-A
  SPAC22H12.04c-1 - 40S ribosomal protein S3aE-B

'SPAC23C4.15':
  rpb5 - DNA-directed RNA polymerases I, II, and III subunit RPABC1; DNA-dependent RNA polymerase catalyzes the transcription of DNA into RNA using the four ribonucleoside triphosphates as substrates. Common component of RNA polymerases I, II and III which synthesize ribosomal RNA precursors, mRNA precursors and many functional non-coding RNAs, and small RNAs, such as 5S rRNA and tRNAs, respectively. Pol II is the central component of the basal RNA polymerase II transcription machinery. Pols are composed of mobile elements that move relative to each other. In Pol II, RPB5 is part of the lower j [...]

'SPACUNK4.06c':

rpb7 - DNA-directed RNA polymerase II subunit rpb7; DNA-dependent RNA polymerase catalyzes the transcription of DNA into RNA using the four ribonucleoside triphosphates as substrates. Component of RNA polymerase II which synthesizes mRNA precursors and many functional non-coding RNAs. Pol II is the central component of the basal RNA polymerase II transcription machinery. It is composed of mobile elements that move relative to each other. RPB7 is part of a subcomplex with RPB4 that binds to a pocket formed by RPB1, RPB2 and RPB6 at the base of the clamp element. The RBP4-RPB7 subcomplex seems [...]

'SPAC6F12.15c':

cut9 - Anaphase-promoting complex subunit cut9; Component of the anaphase-promoting complex/cyclosome (APC/C), a cell cycle-regulated E3 ubiquitin-protein ligase complex that controls progression through mitosis and the G1 phase of the cell cycle. The APC/C is thought to confer substrate specificity and, in the presence of ubiquitin-conjugating E2 enzymes, it catalyzes the formation of protein-ubiquitin conjugates that are subsequently degraded by the 26S proteasome. May play a pivotal role in the control of anaphase

'SPAC2C4.03c':

smd2 - Probable small nuclear ribonucleoprotein Sm D2; Required for pre-mRNA splicing. Required for snRNP biogenesis (By similarity)

'SPBC19G7.05c':

bgs1 - 1,3-beta-glucan synthase component bgs1; Required for the assembly of the division setum and maintenance of cell polarity

'SPAC9G1.02':

wis4 - MAP kinase kinase kinase wis4; Involved in a signal transduction pathway that is activated in under conditions of heat shock, oxidative stress or limited nutrition. Unlike win1, it is not activated by changes in the osmolarity of the extracellular environment. Activates the wis1 MAP kinase kinase by phosphorylation

'SPAC18B11.10':

tup11 - Transcriptional repressor tup11; Transcriptional repressor

Corresponding BC values:

|  | BC | BC stress (60) | BC stress (240) | BC stress 60/BC |
|---|---|---|---|---|
| SPCC16C4.18c | 0.00 | 0.02 | 0.00 | 904.00 |
| SPBC4B4.03 | 5.9e-05 | 0.01 | 0.00 | 100.64 |
| SPBC3E7.14 | 0.00 | 0.04 | 0.01 | 26.98 |
| SPBC16C6.07c | 5.4e-06 | 0.00 | 0.00 | 22.32 |
| SPBC4.07c | 0.01 | 0.08 | 0.05 | 16.03 |
| SPBC15D4.14 | 0.00 | 0.03 | 0.00 | 13.35 |
| SPBC839.10 | 0.01 | 0.08 | 0.01 | 8.29 |
| SPBC32F12.11 | 0.02 | 0.15 | 0.16 | 6.45 |
| SPBC16A3.15c | 0.00 | 0.01 | 0.00 | 5.81 |
| SPBC409.07c | 0.00 | 0.01 | 0.02 | 5.52 |
| SPAC19A8.13 | 0.02 | 0.09 | 0.03 | 5.02 |
| SPBC16E9.12c | 6.3e-06 | 3.1e-05 | 0.00 | 4.92 |
| SPAC22H12.04c | 0.01 | 0.06 | 0.00 | 4.83 |
| SPAC23C4.15 | 0.00 | 0.00 | 0.00 | 4.72 |
| SPACUNK4.06c | 0.04 | 0.16 | 0.18 | 4.37 |
| SPAC6F12.15c | 0.00 | 0.01 | 0.00 | 4.08 |
| SPAC2C4.03c | 0.01 | 0.05 | 0.07 | 4.01 |
| SPBC19G7.05c | 0.00 | 0.02 | 0.02 | 3.89 |
| SPAC9G1.02 | 0.00 | 0.01 | 0.02 | 3.80 |
| SPAC18B11.10 | 0.01 | 0.02 | 0.02 | 3.69 |

*Figure S4*: Network of predicted and known (coloured) interactions for the top 20 genes increasing BC upon stress-treatment. Note the connections between the different proteins in the set (red). From www.bahlerlab.info/PInt, see website for more details.

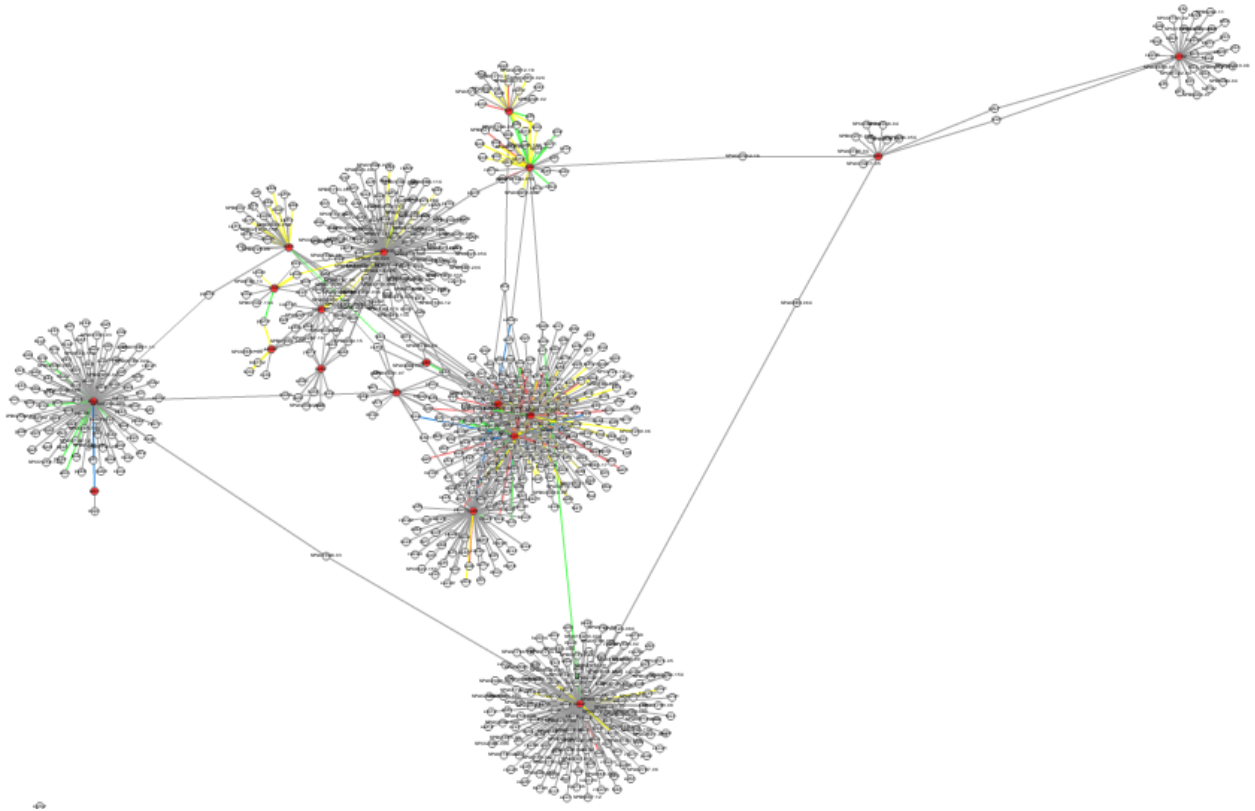

**Supplementary table S7**: *Genes that have higher BC at the late time-point(240') compared to 60' (Top 20 genes ranked by fold change)*

'SPBC16A3.05c':
 rae1 - Poly(A)+ RNA export protein; Required for mitotic cell growth as well as for spore germination. Functions in cell cycle progression through trafficking of proteins required for mitosis. Has a role in the mRNA export process

'SPBC646.09c':
 int6 - Eukaryotic translation initiation factor 3 subunit E; Component of the eIF-3 complex, which binds to the 40S ribosome and promotes the binding of methionyl-tRNAi and mRNA. Required for maintaining the basal level of atf1 and for transcriptional activation of core environmental stress response genes (CESR genes) in response to histidine starvation. May positively regulate proteasome activity. Required for nuclear localization of the proteasome subunit rpn501/rpn502

'SPBC17D11.05':
 tif32 - Eukaryotic translation initiation factor 3 subunit A; Component of the eukaryotic translation initiation factor 3 (eIF-3) complex, which is involved in protein synthesis and, together with other initiation factors, stimulates binding of mRNA and methionyl-tRNAi to the 40S ribosome

'SPAC23E2.01':
 fep1 - Iron-sensing transcription factor 1; Represses the expression of the iron transporter fio1 in response to high iron concentrations. Binds to the consensus sequence 5'-[AT]GATAA-3'. Also represses the expression of str1, str2 and str3

'SPBC14F5.04c':
 pgk1 - Phosphoglycerate kinase

'SPCC1795.11':
 ded1 - ATP-dependent RNA helicase ded1; ATP-binding RNA helicase involved in translation initiation. Remodels RNA in response to ADP and ATP concentrations by facilitating disruption, but also formation of RNA duplexes (By similarity). Inactivation of ded1 blocks mitotic cell cycle progression at G1 and G2/M. Induces sexual development and ascus formation

'SPAC4F10.11':
 spn1 - Septin homolog spn1; Plays a role in the cell cycle. Involved in a late stage of septum formation leading to the separation of the daughter cells

'SPBC409.07c':
 wis1 - Protein kinase wis1; Dosage-dependent regulator of mitosis with serine/ threonine protein kinase activity. May play a role in the integration of nutritional sensing with the control over entry into mitosis. It may interact with cdc25, wee1 and win1. May activate sty1

'SPAC29A4.08c':
 cwf8 - Cell cycle control protein cwf8; Involved in mRNA splicing where it associates with cdc5 and the other cwf proteins as part of the spliceosome

'SPAC9G1.02':
 wis4 - MAP kinase kinase kinase wis4; Involved in a signal transduction pathway that is activated in under conditions of heat shock, oxidative stress or limited nutrition. Unlike win1, it is not activated by changes in the osmolarity of the extracellular environment. Activates the wis1 MAP kinase kinase by phosphorylation

'SPCC18B5.11c':
 Cds1 - Serine/threonine-protein kinase cds1; Has a role in the DNA replication-monitoring S/G2 checkpoint system. It is responsible for blocking mitosis in the S phase. It monitors DNA synthesis by interacting with DNA polymerase alpha and sends a signal to block the onset of mitosis while DNA synthesis is in progress. Phosphorylates rad60

'SPAC22E12.07':
 rna1 - Ran GTPase-activating protein 1; GTPase activator for the nuclear Ras-related regulatory protein spi1 (Ran), converting it to the putatively inactive GDP- bound state

'SPCC576.03c':
 tpx1 - Peroxiredoxin tpx1; Physiologically important antioxidant which constitutes an enzymatic defense against sulfur-containing radicals. Can provide protection against a thiol-containing oxidation system but not against an oxidation system without thiol. Required for the peroxide-induced activation of pap1 via its oxidation and for the nuclear accumulation of pap1. Required also for activation of sty1. Reduced by srx1 and this regulation acts as a molecular switch controlling the transcriptional response to hydrogen peroxide

'SPCC622.09':

htb1 - Histone H2B-alpha; Core component of nucleosome. Nucleosomes wrap and compact DNA into chromatin, limiting DNA accessibility to the cellular machineries which require DNA as a template. Histones thereby play a central role in transcription regulation, DNA repair, DNA replication and chromosomal stability. DNA accessibility is regulated via a complex set of post-translational modifications of histones, also called histone code, and nucleosome remodeling

'SPAC13G7.08c':

crb3 - Pre-rRNA-processing protein crb3/ipi3; Involved in the processing of ITS2 sequences from 35S pre-rRNA (By similarity)

'SPAC20G8.05c':

cdc15 - Cell division control protein 15; After the onset of mitosis, forms a ring-like structure which colocalizes with the medial actin ring. Appears to mediate cytoskeletal rearrangements required for cytokinesis. Essential for viability

'SPAC11H11.06':

arp2 - Actin-related protein 2; Functions as ATP-binding component of the Arp2/3 complex which is involved in regulation of actin polymerization and together with an activating nucleation-promoting factor (NPF) mediates the formation of branched actin networks. Seems to contact the pointed end of the daughter actin filament (By similarity). During cytokinesis it colocalizes to the cortical actin patches until spetation is complete. Has a role in the mobility of these patches. Essential for viability

'SPAC3C7.14c':

obr1 - P25 protein; Unknown. Target of pap1 transcription factor. Confers brefeldin A resistance in S.pombe

'SPBC800.05c':

tub1 - Tubulin alpha-2 chain; Tubulin is the major constituent of microtubules. It binds two moles of GTP, one at an exchangeable site on the beta chain and one at a non-exchangeable site on the alpha-chain

'SPAC2C4.03c':

smd2 - Probable small nuclear ribonucleoprotein Sm D2; Required for pre-mRNA splicing. Required for snRNP biogenesis (By similarity)

Corresponding BC values:

|  | BC | BC stress (60) | BC stress (240) | BC stress (60) / BC | BC stress (240) / BC stress (60) |
|---|---|---|---|---|---|
| SPBC16A3.05c | 0.00 | 0.00 | 0.01 | 0.33 | 14.24 |
| SPBC646.09c | 0.01 | 0.01 | 0.02 | 0.41 | 4.95 |
| SPBC17D11.05 | 0.01 | 0.01 | 0.04 | 1.34 | 4.58 |
| SPAC23E2.01 | 0.00 | 0.00 | 0.00 | 0.52 | 3.80 |
| SPBC14F5.04c | 0.01 | 0.00 | 0.01 | 0.27 | 2.41 |
| SPCC1795.11 | 0.08 | 0.07 | 0.14 | 0.79 | 2.15 |
| SPAC4F10.11 | 2.5e-06 | 6.3e-06 | 1.2e-05 | 2.52 | 1.90 |
| SPBC409.07c | 0.00 | 0.01 | 0.02 | 5.52 | 1.92 |
| SPAC29A4.08c | 0.08 | 0.05 | 0.10 | 0.62 | 1.91 |
| SPAC9G1.02 | 0.00 | 0.01 | 0.02 | 3.80 | 1.84 |
| SPCC18B5.11c | 0.01 | 0.01 | 0.02 | 0.66 | 1.80 |
| SPAC22E12.07 | 0.00 | 0.00 | 0.01 | 1.28 | 1.53 |
| SPCC576.03c | 0.02 | 0.00 | 0.00 | 0.14 | 1.51 |
| SPCC622.09 | 0.00 | 0.01 | 0.01 | 1.58 | 1.48 |
| SPAC13G7.08c | 0.02 | 0.02 | 0.02 | 0.77 | 1.47 |
| SPAC20G8.05c | 0.03 | 0.02 | 0.03 | 0.84 | 1.47 |
| SPAC11H11.06 | 0.00 | 0.00 | 0.00 | 0.86 | 1.38 |
| SPAC3C7.14c | 0.01 | 0.01 | 0.01 | 1.35 | 1.34 |
| SPBC800.05c | 0.01 | 0.01 | 0.01 | 0.65 | 1.34 |
| SPAC2C4.03c | 0.01 | 0.05 | 0.07 | 4.01 | 1.31 |

*Figure S5:* Network of predicted and known (coloured) interactions for the top 20 genes increasing BC in late stress compared to stress. Note the presence of many interactions interconnecting the proteins in the set (red). From www.bahlerlab.info/PInt, see website for more details.

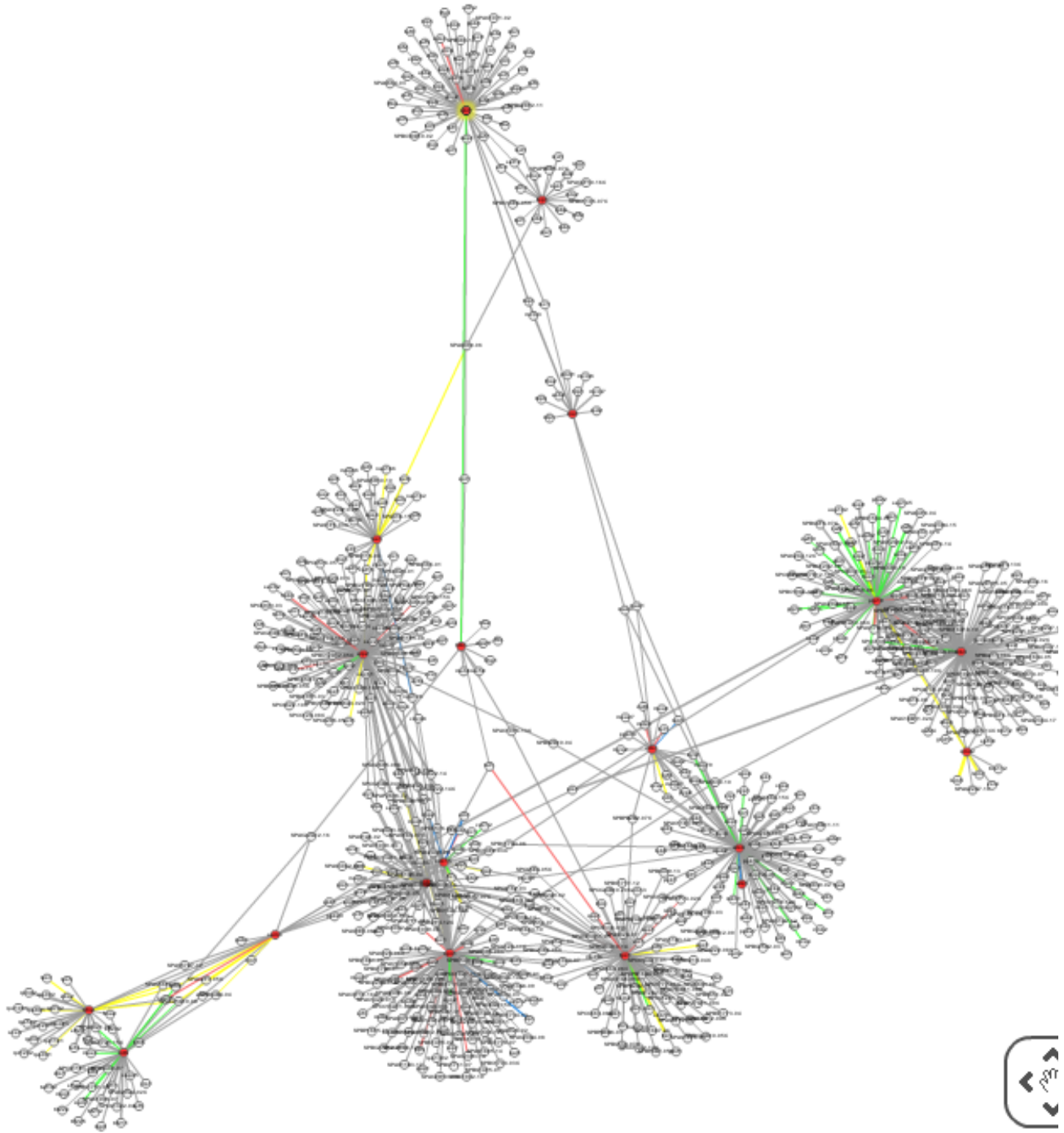